\documentclass[english,12pt,a4paper]{article}
\usepackage{authblk}
\usepackage[text={17.0cm,23.7cm}]{geometry}
\usepackage[T1]{fontenc}
\usepackage[utf8]{inputenc}
\usepackage[english]{babel} 
\usepackage{hyperref}
\usepackage{graphicx}
\usepackage{amsmath}
\usepackage{amssymb}
\usepackage[dvipsnames]{xcolor}
\usepackage{yfonts}
\usepackage[mathscr]{euscript}
\usepackage[style=numeric-comp, sorting=none]{biblatex}
\usepackage{float}
\usepackage{csquotes}
\usepackage[normalem]{ulem}

\usepackage{bm}
\usepackage{bbold}
\usepackage{relsize}
\usepackage{caption}
\usepackage{subcaption}

\begin{document}
	
	\title{\vspace{-2cm}
		{\normalsize \begin{flushright}
            KEK-TH-2788\\
            YITP-25-191 
            \end{flushright}
            }
		\vspace{0.6cm}
	\textbf{Quantum effects on neutrino parameters from a flavored gauge boson}\\[8mm]}
\author[1]{Alejandro Ibarra}
\author[2, 3, 4, 1]{Lukas Treuer}
\affil[1]{\normalsize\textit{Technical University of Munich, TUM School of Natural Sciences, Physics Department, James-Franck-Strasse 1, 85748 Garching, Germany}}
\affil[2]{\normalsize\textit{Yukawa Institute for Theoretical Physics, Kyoto University, Kitashirakawa Oiwakecho, Kyoto 606-8502, Japan}}
\affil[3]{\normalsize\textit{KEK Theory Center, 1-1 Oho, Tsukuba 305-0801, Japan}}
\affil[4]{\normalsize\textit{The Graduate University for Advanced Studies (SOKENDAI), 1-1 Oho, Tsukuba 305-0801, Japan}}

\date{}
\maketitle

\begin{abstract}

We calculate the one-loop renormalization group equations of the neutrino mass matrix when the Standard Model particle content is extended with a massive gauge boson which has family-dependent couplings to the left-handed leptons. We show that quantum effects induced by the extra gauge boson increase the rank of the neutrino mass matrix at the one-loop level, in contrast to the well-known result that Standard Model fields can only increase the rank at the two-loop level. We also discuss the possibility of generating dynamically the measured mass differences and mixing angles between the active neutrinos in scenarios with normal and inverted mass ordering.

\end{abstract}
\section{Introduction}
\label{sec:Intro}

In the simplest models of neutrino mass generation, the smallness of neutrino masses is associated to the breaking of lepton number by two units by very heavy particles. Below the energy scale of the lightest particle with lepton-number-breaking interactions, the model can be conveniently described by the Standard Model extended by a dimension-5 lepton-number-breaking operator, 
commonly called the Weinberg operator~\cite{Weinberg:1979sa},  with a Wilson coefficient determined by the masses of the new particles and their couplings to the left-handed leptons and Higgs doublet. 
After electroweak symmetry breaking, this term gives rise to Majorana masses for the left-handed neutrinos, which are naturally small if the masses of the new particles are much larger than the scale of electroweak symmetry breaking.

Quantum effects can modify the size and flavor structure of the Wilson coefficient of the Weinberg operator. In view of the postulated large hierarchy between the cut-off scale of the effective field theory and the scale at which the neutrino masses and mixing angles are measured, the largest quantum effects are encapsulated in the Renormalization Group Equations (RGEs). The one-loop RGE of the Weinberg operator in the Standard Model was first calculated in refs.~\cite{Chankowski:1993tx,Babu:1993qv,Antusch:2001ck} while the complete two-loop RGE was first calculated in \cite{Ibarra:2024tpt} (partial results were presented in ~\cite{Davidson:2006tg}). The phenomenological implications have been discussed, e.g., in ~\cite{Casas:1999tp,Ellis:1999my,Casas:1999kc,Casas:1999tg,Chankowski:1999xc,Haba:1999ca,Haba:1999ca,Chankowski:2000fp,Chankowski:2001mx}, demonstrating that one-loop quantum effects can have a strong impact on scenarios with degenerate neutrinos, leading to quasi-fixed points in the infrared for the elements of the leptonic mixing matrix, and generating a mass splitting between the initially degenerate eigenvalues.  Furthermore, two-loop effects can have a strong impact on scenarios with very hierarchical neutrinos, softening mass hierarchies which are at the cut-off scale much larger than $(16\pi^2)^2$. In particular, two-loop effects can increase the rank of the mass matrix.

It is plausible that there could be new degrees of freedom in the ``desert'' between the electroweak scale and the scale of decoupling of, e.g., the right-handed neutrinos. A renowned example is supersymmetry, which also ensures the stability of the Higgs boson mass under large quantum effects induced by the heavy right-handed neutrinos. The one-loop RGE of the Weinberg operator in the Minimal Supersymmetric Standard Model has been studied in \cite{Babu:1993qv,Chankowski:1993tx,Casas:1999ac,Antusch:2001vn}, and the two-loop RGE in \cite{Antusch:2002ek}. Another example is the extension of the Standard Model by a second Higgs doublet. The one-loop RGE of the Weinberg operator in this scenario has been studied in \cite{Babu:1993qv,Chankowski:1993tx,Grimus:2002nk,Grimus:1999wm,Antusch:2001vn,Ibarra:2011gn}.

In this work, we consider the effects on the Weinberg operator in gauge extensions of the Standard Model, and specifically in those that act differently on different fermion generations. We will consider for concreteness a spontaneously broken $\textrm{U}(1)_{L_\mu - L_\tau}$ symmetry \cite{He:1990pn,Heeck:2011wj}, where the corresponding massive $Z'$ gauge boson is lighter than the cut-off scale up to which the effective theory with a Weinberg operator remains valid. We will show that in this extension the RGE effects on the Weinberg operator are qualitatively different than in the effective theory containing just the Standard Model fields. Concretely, we will show that the rank of the Majorana neutrino mass matrix can be raised at the one loop level, which is the main new result of this paper, and we will discuss some phenomenological implications of the novel structure of the RGEs. 

This paper is organized as follows. In section
\ref{sec:Model} we discuss a renormalizable scenario with right-handed neutrinos and $\textrm{U}(1)_{L_\mu-L_\tau}$ symmetry, which generates an effective theory containing the Weinberg operator and a massive flavored gauge boson. This provides a concrete example for a renormalizable model admitting the phenomena discussed thereafter. In section 
\ref{sec:RGE} we study the RGE of the Weinberg operator and discuss some of its implications, and finally in section \ref{sec:conclusions} we present our conclusions.

\section{A scenario with Majorana neutrinos and a flavored \texorpdfstring{$Z'$}{Z'}} 
\label{sec:Model}

We extend the Standard Model gauge group by an Abelian $\textrm{U}(1)$ symmetry that acts differently on the left-handed leptons of different generations. In this paper, we will consider a local $\textrm{U}(1)_{L_\mu-L_\tau}$ symmetry as archetypal example of family-dependent gauge symmetry, although our analysis can be straight-forwardly generalized to others. The gauge symmetry leads to interactions of the gauge boson $Z'$ with the left-handed leptons of the form:
\begin{equation}
	{\cal L} \supset - g' Q'_{fg} \overline{l_f} \gamma^\mu Z^\prime_\mu l_g,
    \label{eq:L-gauge}
\end{equation}
where $g'$ is the coupling constant, $l_f$ with $f=e,\mu,\tau$ denotes the lepton doublets, and $Q'_{fg}$ is a matrix that for the $\textrm{U}(1)_{L_\mu-L_\tau}$ symmetry  reads:
\begin{align}
	Q' = \begin{pmatrix}
		0 & 0 & 0\\
		0 & 1 & 0\\
		0 & 0 & -1
	\end{pmatrix},
\label{eq:Q-def}
\end{align}
and contains the $\textrm{U}(1)_{L_\mu-L_\tau}$ charges of the leptons.

The gauge symmetry of the model permits a dimension-5 Weinberg operator, which has the form 
\begin{align}
\label{eq:WeinbergOperator}
    \mathcal{L}_{int} &\supset \frac{1}{4}\, \kappa_{g f}\,\overline{l_c^{g,\, C}}\, \varepsilon^{c d}\, \phi_d\, l_b^f\, \varepsilon^{b a}\, \phi_a\, + \mathrm{h.c.} \\
\label{eq:WeinbergOperatorFlavorStructure}
    &\sim \kappa_{g f}\, l^g\, l^f\, \phi\, \phi\, + \mathrm{h.c.},
\end{align}
with $l^C = i \gamma^2 \gamma^0 l^*$ being the conjugate lepton doublet spinor; $\phi$, the Higgs doublet; $g$ and $f$, flavor indices; $a,\, b,\, c,\,$ and $d$, $\textrm{SU}(2)_L$ indices; $\varepsilon$, the totally antisymmetric tensor; and $\kappa$, the corresponding Wilson coefficient. In general, this operator will give rise to Majorana masses for the left-handed neutrinos after electroweak symmetry breaking, where the Higgs field obtains a non-zero vacuum expectation value.

If the $\textrm{U}(1)_{L_\mu-L_\tau}$ symmetry is exact in the model, the Wilson coefficient takes the well-known form 
\begin{align}
	\kappa_{gf}\Big|_{\rm unbroken}=\begin{pmatrix} 
		\kappa_{11} & 0 & 0 \\
		0 & 0 &\kappa_{23}  \\
		0  & \kappa_{23} &0
		\end{pmatrix},
        \label{eq:Weinberg_unbroken}
\end{align}
which is preserved at any scale, due to the invariance of the action under  $\textrm{U}(1)_{L_\mu-L_\tau}$. However, this structure is excluded, e.g., by neutrino oscillation data \cite{Esteban:2024eli}.

On the other hand, the $\textrm{U}(1)_{L_\mu-L_\tau}$ symmetry could be broken at a high-energy scale, which will reflect on the Wilson coefficient $\kappa$. In the following, we consider an extension of the model with three right-handed neutrinos with $\textrm{U}(1)_{L_\mu - L_\tau}$ charges equal to 0, 1, and $-1$, thus ensuring the cancellation of gauge anomalies. Further, we introduce complex scalar fields, also singlets under the Standard Model gauge group, and with $\textrm{U}(1)_{L_\mu - L_\tau}$ charges equal to 0, $-1$, and $-2$. The additional field content and charges are summarized in table~\ref{tab:add_rh_neutrinos_scalars}.

\begin{table}[t!]
    \centering
    \begin{tabular}{c|c||c|c}
       Field & $\textrm{U}(1)_{L_\mu - L_\tau}$ charge & Field & $\textrm{U}(1)_{L_\mu - L_\tau}$ charge \\
    \hline
       $N_1$ & $0$ & $S_0$ & $0$ \\
       $N_2$ & $1$ &  $S_1$ & $-1$\\
       $N_3$ & $-1$ &  $S_2$ & $-2$ 
    \end{tabular}
    \caption{Additional right-handed neutrinos and scalar fields charged under $\textrm{U}(1)_{L_\mu - L_\tau}$, and neutral under the Standard Model.}
    \label{tab:add_rh_neutrinos_scalars}
\end{table}

With this particle content, the part of the Lagrangian involving the new fermionic mass terms and Yukawa couplings reads:
\begin{align}
2{\cal L}_{\rm mass} &=  M_{11} \overline{N_1^C} N_1 +  M_{23} \overline{N_2^C} N_3 + {\rm h.c.}, \nonumber \\[5pt]
2{\cal L}_{\rm Yuk} &= 2 Y_{\nu, e 1} \bar l_e \epsilon \phi^* N_1 + 2 Y_{\nu, \mu 2} \bar l_\mu \epsilon \phi^* N_2 + 2Y_{\nu,\, \tau 3} \overline{l_\tau} \epsilon \phi^* N_3 + \lambda_{11} S_0 \overline{N_1^C} N_1 + \lambda_{23} S_0 \overline{N_2^C} N_3 + \nonumber \\
&\quad + \lambda_{12} S_1 \overline{N_1^C} N_2 + \lambda_{13} S_1^\dagger \overline{N_1^C} N_3 + \lambda_{22} S_2 \overline{N_2^C} N_2 +  \lambda_{33} S_2^\dagger \overline{N_3^C} N_3 + {\rm h.c.},
\end{align}
where we assume all couplings to be real for simplicity. In general, all scalars will acquire non-zero vacuum expectation values, that we also assume to be real. Then, the right-handed neutrino mass matrix reads:
\begin{equation}
	M_R = \begin{pmatrix}
		M_{11} +\lambda_{11}\langle S_0\rangle & \lambda_{12}\, \langle S_1\rangle  & \lambda_{13}\, \langle S_1\rangle \\
		\lambda_{12}\, \langle S_1\rangle & \lambda_{22}\, \langle S_2\rangle & M_{23}+\lambda_{23}\langle S_0\rangle \\
		\lambda_{13}\, \langle S_1\rangle  & M_{23}+\lambda_{23}\langle S_0\rangle & \lambda_{33}\, \langle S_2\rangle
	\end{pmatrix}.
	\label{eq:MassMatrixRHNeutrino}
\end{equation}

This amounts to the most general form of the right-handed neutrino mass matrix, with eigenvalues $|M_1|\leq |M_2| \leq |M_3|$. Further, the spontaneous breaking of the $\textrm{U}(1)_{L_\mu - L_\tau}$ symmetry generates a mass for the gauge boson $Z'$ given by
\begin{align}
M_{Z'} = \sqrt{2} g' \cdot \sqrt{\langle S_1\rangle^2 + 4\langle S_2\rangle^2}.
\label{eq:MassZPrimeThreeScalar}
\end{align}

Upon integrating out the right-handed neutrinos, one finds at the scale $\mu=M_1$ an effective Weinberg operator given by $\kappa=-Y_\nu M_R^{-1} Y_\nu^T$. If the $\textrm{U}(1)_{L_\mu - L_\tau}$ symmetry is only mildly broken, i.e., when $\langle S_1 \rangle,~\langle S_2 \rangle\ll M_{11},~M_{23}$, 
the  Wilson coefficient for the Weinberg operator takes the form eq.~\eqref{eq:Weinberg_unbroken}, perturbed by terms of order $\langle S_i\rangle/M_{jk}$. Since the $\textrm{U}(1)_{L_\mu - L_\tau}$ symmetry is only mildly broken, the structure eq.~\eqref{eq:Weinberg_unbroken} remains approximate at all orders in perturbation theory, which is incompatible with neutrino oscillation experiments, as previously mentioned.

On the other hand, if the $\textrm{U}(1)_{L_\mu - L_\tau}$ symmetry is badly broken, i.e., when $\langle S_1 \rangle,~\langle S_2 \rangle\gg M_{11},~M_{23}$, it will be possible to reproduce the low energy neutrino data by suitable choices of   $Y_\nu$, $\lambda_{ij}$, $\langle S_0\rangle$, $\langle S_1\rangle$, and $\langle S_2\rangle$. In this case, and since there is no remnant of the $\textrm{U}(1)_{L_\mu-L_\tau}$ symmetry in the neutrino mass matrix, renormalization group effects can markedly modify its flavor structure between the scale $M_1$ and the scale of neutrino experiments. 

The effective theory below $M_1$ may only consist of the Standard Model extended by the Weinberg operator, or may also include the $Z'$ as a dynamical degree of freedom. In the latter case, the $Z'$ will also affect the running of the neutrino mass matrix, which is the focus of this work. 

\begin{figure}[t!]
    \centering
    \includegraphics[width=0.5\textwidth]{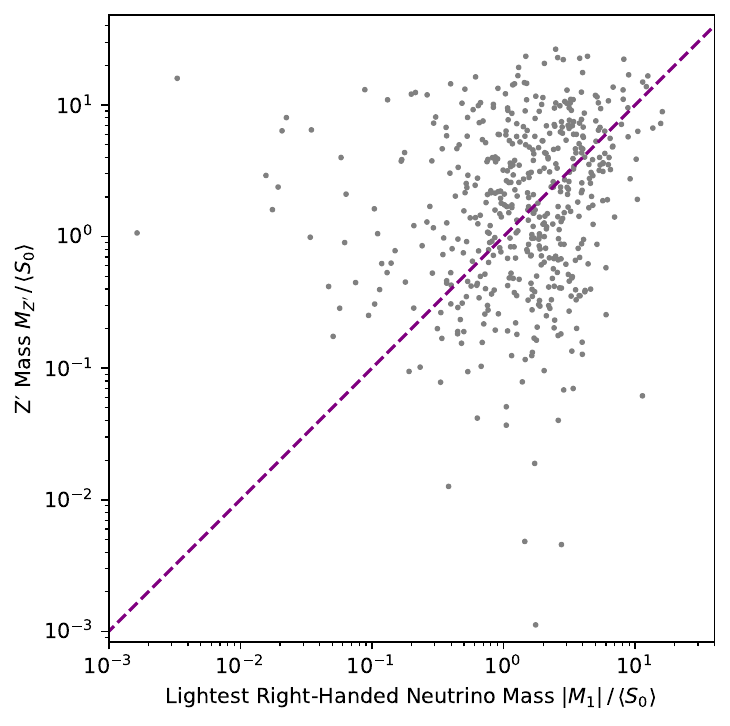}
    \caption{Scatter plot of the absolute value of the smallest right-handed neutrino mass $|M_1|$ vs. $Z'$ gauge boson mass $M_{Z'}$ from random, uniform scans of $\lambda, \langle S_1 \rangle/\langle S_0 \rangle, \langle S_2 \rangle/\langle S_0 \rangle,$ and $g'$. The values admitting the RGE effect discussed here are below the diagonal with $M_{Z'} = |M_1|$. }
    \label{fig:m1_vs_mzprime}
\end{figure}

Let us consider for illustration a scenario with $\langle S_i\rangle \sim S \times {\cal O}(1)$, $\lambda_{ij} \sim \lambda \times {\cal O}(1)$, so that $M_i=\lambda S \times {\cal O}(1)$. Further, let us consider $Y_{\nu, \alpha i}=y_\nu\times{\cal O}(1)$. Then, the flavor structure of $\kappa$ can be adjusted by choosing appropriately the ${\cal O}(1)$ parameters, while the overall size of the neutrino masses is approximately given by $m_\nu= y_\nu^2 \langle H^0\rangle^2/(\lambda S)$. On the other hand, the mass of the $Z'$ is $M^2_{Z'}= g^{\prime 2} S^2 \times {\cal O}(1)$. Therefore, the ratio between the lightest right-handed neutrino mass scale and the $Z'$ mass scale is $M_1/M_{Z'}=\lambda/g' \times {\cal O}(1)$, which can be larger or smaller than 1. We show in figure~\ref{fig:m1_vs_mzprime} a scatter plot confirming this expectation by taking a random, uniform scan of $\lambda_{ij}=[-5,5]$, $\langle S_1 \rangle/\langle S_0 \rangle, \langle S_2 \rangle/\langle S_0 \rangle=10^{[-1, 1]}$ and $g'=[0,1]$. As apparent from the figure, there are numerous points for which $M_{Z'} < |M_1|$ such that the $Z'$ will affect the running of the Weinberg operator at scales $M_{Z'}\leq \mu \leq |M_1|$.

\section{RGE of the Weinberg operator with a \texorpdfstring{$Z'$}{Z'}}
\label{sec:RGE}

\begin{figure}[t!]
    \centering
    \includegraphics[width=0.35\textwidth]{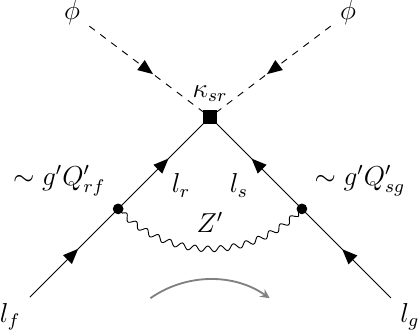}
    \caption{One-loop diagram of $Z'$ giving rise to the $G^T \kappa G$ term in the RGE. The light gray arrow denotes fermion flow used for the calculation, following~\cite{Denner:1992vza}.}
\label{fig:vertex_correction}
\end{figure}

At energy scales $\mu$ between the cut-off scale $M_1$ (or more generally $\Lambda$) and the decoupling scale $M_{Z'}$ of the $Z'$ the one-loop RGE of the Weinberg operator has the form:
\begin{align}
	  16\pi^2\frac{d\kappa}{d t}\, =\, \alpha\, \kappa\, +\, P^T\, \kappa\, +\, \kappa\, P\, +\, G^T\, {\kappa}\, G,
	\label{eq:RGE}
\end{align}
with $t=\log(\mu/M_1)$ and
\begin{align}
\alpha&=  \lambda-3g_2^2 +2 {\rm Tr}( 3Y_u^\dagger Y_u+3Y_d^\dagger Y_d+Y_e^\dagger Y_e), \nonumber \\
P&= -\frac{3}{2}(Y^\dagger_e Y_e), \nonumber \\
G&= \sqrt{6} g' Q'.
\end{align}
Here, $\lambda$ is the quartic coupling in the Higgs potential; $g_2$ is the $\textrm{SU}(2)_L$ gauge coupling constant; $Y_u,~Y_d$, and $Y_e$ are respectively the Yukawa couplings of the up-type quarks, down-type quarks, and charged leptons; and $g'$ and $Q'$ are respectively the gauge coupling constant of the $\textrm{U}(1)_{L_\mu-L_\tau}$ symmetry and the matrix of charges defined in eq.~\eqref{eq:Q-def}. The terms proportional to $\kappa$ and $P$ correspond to the Standard Model contributions, and have been calculated first in \cite{Babu:1993qv, Chankowski:1993tx} and later revised in
\cite{Chankowski:2001hx,Antusch:2001ck}. The term containing the matrix $G$ encodes the quantum effects induced by the $Z'$ interaction with the charged leptons through the diagram shown in figure~\ref{fig:vertex_correction}. Clearly, if the $Z'$ is flavor independent, $G$ is proportional to the identity, and the term $G^T \kappa G$ can be written in the form $\alpha\kappa$.\footnote{Note that the gauge coupling for $\textrm{U}(1)_Y$ does not appear because the divergent contributions to $\kappa$ from the vertex renormalization and the self-energy of the Higgs doublet exactly cancel.}

Notably, the new term in eq.~\eqref{eq:RGE} takes the same form as the two-loop contribution of the Yukawa coupling in the Standard Model \cite{Davidson:2006tg, Ibarra:2024tpt}, which is known to increase the rank of the mass matrix  (an analogous rank-increasing term also exists in the two-loop RGE of the right-handed neutrino mass matrix \cite{Ibarra:2018dib,Ibarra:2020eia}). However, we find that with the flavor-dependent gauge  interactions of the $Z'$, the rank-increasing term appears at the one-loop level. This is the main result of this paper.\footnote{One can explicitly show that at the one-loop level, the rank-increasing term can only appear due to flavor-dependent gauge interactions.}

Considering only the effect of the $\tau$ Yukawa coupling $y_{\tau}$ for simplicity, and using $Q_{ij}'=q'_i\delta_{ij}$ with $q'_i=0,1,-1$ for the electron, muon and tau flavors, eq.~\eqref{eq:RGE} can be analytically solved as
\begin{align}
\kappa_{ij}(t)=\kappa_{ij}(t_{M_1})\exp\Big\{\frac{1}{16\pi^2}\int_{t_\Lambda}^t dt'\Big[\alpha(t')-\frac{3}{2} y_\tau(t')^2(\delta_{i3}+\delta_{3j})+6g'(t')^2q'_i q'_j\Big]\Big\}.
\label{eq:exact}
\end{align}

For the interpretation of the results, it is useful to decompose $\kappa=U^T D_\kappa U$,
where $U$ is the unitary leptonic mixing matrix, and $D_\kappa={\rm diag}(\kappa_1,\kappa_2,\kappa_3)$ is a diagonal matrix~\cite{Casas:1999tg}. The RGEs for $\kappa_i$ and $U$ read:
\begin{align}
    16\pi^2 \frac{{\rm d} \kappa_i}{{\rm d} t} = \alpha \kappa_i + 2 \tilde{P}_{ii} \kappa_i + {\rm Re}\sum_{j = 1}^3 \big[ \tilde{G}_{ji} \big]^2 \kappa_j,
    \label{eq:RGE_Eigenvalues}
\end{align}
and 
\begin{equation}
    \frac{{\rm d} U}{{\rm d} t} = T\, U,
    \label{eq:U_RGE}
\end{equation}
with
\begin{equation}
    16 \pi^2 T_{ij} = \begin{cases}
    \begin{aligned}
    i \sum_{k = 1}^n {\rm Im}\big[\tilde{G}_{ki}^2\big] \frac{\kappa_k}{2\, \kappa_i} \end{aligned} 
    &\text{for $i = j$,}
    \\[15pt]
    \begin{aligned}
    &\frac{\kappa_i + \kappa_j}{\kappa_i - \kappa_j} {\rm Re} [\tilde{P}_{ij}] +  \sum_{k = 1}^3 \frac{\kappa_k}{\kappa_i - \kappa_j} {\rm Re} \big[\tilde{G}_{ki}\, \tilde{G}_{kj}\big]\,   \\[5pt]
    &+ i \bigg(\frac{\kappa_i- \kappa_j}{\kappa_i + \kappa_j} {\rm Im} \big[\tilde{P}_{ij}\big] + \sum_{k = 1}^3  \frac{\kappa_k}{\kappa_i + \kappa_j}{\rm Im} \big[\tilde{G}_{ki}\, \tilde{G}_{kj}\big]\, \bigg)
    \end{aligned} 
    &\text{for $i \neq j$,}
    \end{cases}
    \label{eq:T_RGE}
\end{equation}
where $\tilde{P}=U P U^\dagger$ and $\tilde{G}=U G U^\dagger$. The results for non-abelian flavor-gauge theories are given in appendix~\ref{appendix:non-abelian}.

As apparent from  eq.~\eqref{eq:RGE_Eigenvalues}, the RGE for $\kappa_i$ is not necessarily proportional to $\kappa_i$ itself (as long as $\tilde G_{ji}\neq 0$ for $i\neq j$), which results in the increase of the rank of the mass matrix, or in a potentially large quantum effect in the lightest neutrino mass when $M_{Z'}< M_1$.

\subsection{Two-generation case}

To illustrate the novel effect induced by the $Z'$ let us consider a simplified scenario with only two generations, and where all parameters are real.
We denote the eigenvalues of the matrix $\kappa$ at the cut-off $M_1$ as $\kappa_1$, $\kappa_2$, and the leptonic mixing matrix as
\begin{align}
U=\begin{pmatrix}
\cos\theta &\sin\theta \\
-\sin\theta & \cos\theta
\end{pmatrix},
\end{align}
so that
\begin{align}
\widetilde G= \sqrt{6} g' \begin{pmatrix} \cos2\theta &-\sin2\theta  \\ -\sin2\theta & -\cos2\theta
\end{pmatrix},
\end{align}
where we used $G = \sqrt{6} g' {\rm diag}(1, -1)$. Neglecting the effect of the charged lepton Yukawa couplings (encoded in $\tilde P$), one finds the approximate solutions
\begin{align}
\kappa_1|_{M_{Z'}}&\simeq (a - b\cos^2 2\theta) \kappa_1 - b\sin^2 2\theta  \kappa_2, \\
\kappa_2|_{M_{Z'}}&\simeq - b\sin^2 2\theta \kappa_1+(a - b\cos^2 2\theta) \kappa_2,
\label{eq:mass-2gen}
\end{align}
where $a=1 - \alpha/(16\pi^2)\log (M_1/M_{Z'})$, $b=6 g^{\prime 2}/(16\pi^2)\log (M_1/M_{Z'})$.

From these expressions, it is apparent that even if $\kappa_1$ vanishes at $M_1$, quantum effects will generate a non-zero value at $M_{Z'}$. More concretely, while the mass hierarchy is $\kappa_1/\kappa_2=0$ at the cut-off scale $M_1$ (or more generally $\Lambda$), the mass hierarchy at the scale $M_{Z'}$ is
\begin{align}
\frac{\kappa_1}{\kappa_2}\Big|_{M_{Z'}} \simeq \frac{-b \sin^22\theta }{a-b\cos^2 2\theta} \simeq - \frac{6 g^{\prime 2}}{16\pi^2}\log\left(\frac{M_1}{M_{Z'}}\right)\sin^2 2\theta,
\end{align}
which typically gives a hierarchy $\sim{\cal O}(10^{-2})$ for $g'\sim 1$. 

On the other hand, if the two eigenvalues are degenerate at the cut-off scale, then at the scale $M_{Z'}$
\begin{align}
\frac{\kappa_1}{\kappa_2}\Big|_{M_{Z'}}=1,
\end{align}
i.e., they remain degenerate (up to effects induced by charged lepton Yukawa couplings).

\subsection{Three-generation case}

For the three-generation case, and again assuming real parameters, the matrix $\tilde G$ can be cast as $\tilde G=\sqrt{6} g' W$, where 
\begin{align}
W=\begin{pmatrix}
U_{12}^2-U_{13}^2 & U_{12}U_{22} -U_{13}U_{23}  & U_{12}U_{32} -U_{13} U_{33}  \\
U_{22} U_{12} -U_{23}U_{13}  & U_{22}^2-U_{23}^2 & U_{22}U_{32} -U_{23}U_{33}  \\
U_{32} U_{12} -U_{33}U_{13}  & U_{32}U_{22} -U_{33}U_{23}  & U_{32}^2-U_{33}^2
\end{pmatrix}.
\end{align}

Neglecting the effects from Yukawa couplings, one obtains as eigenvalues of $\kappa$ at the scale $M_{Z'}$:
\begin{align}
{\rm eig}\big(\kappa|_{M_{Z'}}\big)\simeq
\Big\{&\big(a - b W_{11}^2\big)\kappa_1- b W_{21}^2  \kappa_2 - b W_{31}^2  \kappa_3,
\nonumber \\
&-b W_{12}^2 \kappa_1+ \big(a - b W_{22}^2\big)\kappa_2- b W_{32}^2  \kappa_3,
\nonumber \\
&-b W_{13}^2\kappa_1- b W_{23}^2\kappa_2+ \big(a - b W_{33}^2\big) \kappa_3\Big\},
\label{eq:mass-3gen}
\end{align}
where $a$ and $b$ were defined after eq.~\eqref{eq:mass-2gen}.

The expected neutrino mass ordering depends chiefly on the ordering of the moduli of the eigenvalues of $\kappa$ at the cut-off scale. Let us consider two different scenarios: $|\kappa_1|, |\kappa_2| \ll |\kappa_3|$, which naturally leads to a normal ordering at low energies, and $|\kappa_3|\ll |\kappa_1|, |\kappa_2|$, which naturally leads to an inverted ordering at low energies.

\subsubsection{Normal mass hierarchy: \texorpdfstring{$|\kappa_1|,\, |\kappa_2|\ll |\kappa_3|$}{|k1|, |k2| << |k3|}}

One of the most notable features of the scenario with a flavored gauge symmetry is that the lightest neutrino mass can receive at the one-loop level significant contributions from the heavier masses. Let us consider the limiting case where $\kappa_1=0$ and $\kappa_2<\kappa_3$ at the cut-off scale. Then, at the scale $M_{Z'}$, one finds a spectrum with normal ordering~\footnote{Note that depending on the ratio $\kappa_2/\kappa_3$ at the cut-off scale, we may also obtain inverted ordering.} with
\begin{align}
\kappa_1\big|_{M_{Z'}}&\simeq - b W_{31}^2 \kappa_3 \nonumber \\
\kappa_2\big|_{M_{Z'}}&\simeq \kappa_2 \nonumber \\
\kappa_3\big|_{M_{Z'}}&\simeq \kappa_3,
\label{eq:NMH-k1eq0}
\end{align}
which gives a hierarchy between the lightest and the heaviest neutrino given by:
\begin{align}
\frac{\kappa_1}{\kappa_3}\Big|_{M_{Z'}}
 \simeq -\frac{6 g'^2}{16\pi^2}
(U_{32}U_{12}- U_{33}U_{13})^2\log\left(\frac{M_1}{M_{Z'}}\right).
\end{align}

The same conclusion holds even when $\kappa_1\neq 0$ at the cut-off scale, as long as the hierarchy $(\kappa_1/\kappa_3)\big|_{M_1}$ is smaller than the one generated by quantum effects.

If two eigenvalues are degenerate at $M_1$, $\kappa_i=\kappa_j$, then the leptonic mixing matrix reaches a quasifixed value in the infrared, corresponding to 

\begin{equation}
    \sum_k \tilde G_{ki} \tilde G_{kj} \kappa_k=0,
    \label{eq:degen_quasifixed_val}
\end{equation}
as follows from eq.~\eqref{eq:T_RGE}. For instance, if $\kappa_1=\kappa_2$  and $|\kappa_2|\ll |\kappa_3|$, the quasifixed value is $\tilde G_{31} \tilde G_{32} \kappa_3\simeq 0$ (or equivalently $W_{31} W_{32}\simeq 0$), so that the masses at low energies read:
\begin{align}
\big|\kappa_1|_{M_{Z'}}\big|&\simeq {\min_{i=1,2}}\Big\{\big|
\big(a-b W_{1i}^2-b W_{2i}^2\big)\kappa_2-b W_{3i}^2 \kappa_3\big|\Big\},
\nonumber \\
\big|\kappa_2|_{M_{Z'}}\big|&\simeq {\max_{i=1,2}}\Big\{
\big|
\big(a-b W_{1i}^2-b W_{2i}^2\big)\kappa_2-b W_{3i}^2 \kappa_3\big|\Big\},\nonumber \\
\big|\kappa_3|_{M_{Z'}}\big|&\simeq  |\kappa_3|.
\label{eq:mass-3gen-normal}
\end{align}

Neglecting the terms $\propto \kappa_2$, the quasifixed value is $W_{31} \simeq 0$ such that the  leptonic mixing angles satisfy:\footnote{In our convention $\kappa = U^T D_\kappa U$ leading to eq.~\eqref{eq:angle-3gen-normal}, there are no solutions for $0 < \theta_{ij} < 90^\circ$ with vanishing CP-phase $\delta_{\rm CP} = 0$. However, taking $\delta_{\rm CP}=180^\circ$ or rephasing the eigenvector $\nu_3 \rightarrow -\nu_3$ admits solutions due to an overall sign-flip in eq.~\eqref{eq:angle-3gen-normal}. To avoid introducing additional parameters at this point for clarity, we instead equivalently take $-90^\circ < \theta_{13} < 0$ here. Choosing a different convention such as $\kappa = U^* D_\kappa U^\dagger$ may give quasifixed points with different relations between the mixing angles.}
\begin{align}
\sin\theta_{13} \simeq -\frac{\sin\theta_{12}\cos\theta_{12}\tan\theta_{23}}{1+\sin^2\theta_{12}}.
\label{eq:angle-3gen-normal}
\end{align}

\begin{figure}[t!]
    \centering
    \includegraphics[width=0.45\textwidth]{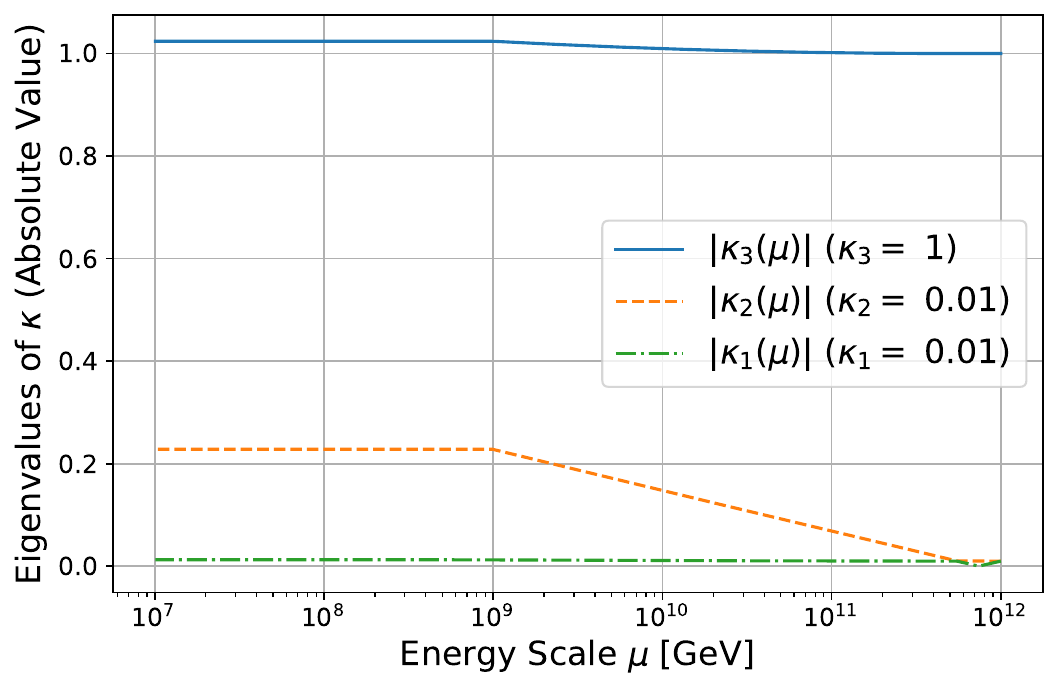}~~
    \includegraphics[width=0.45\textwidth]{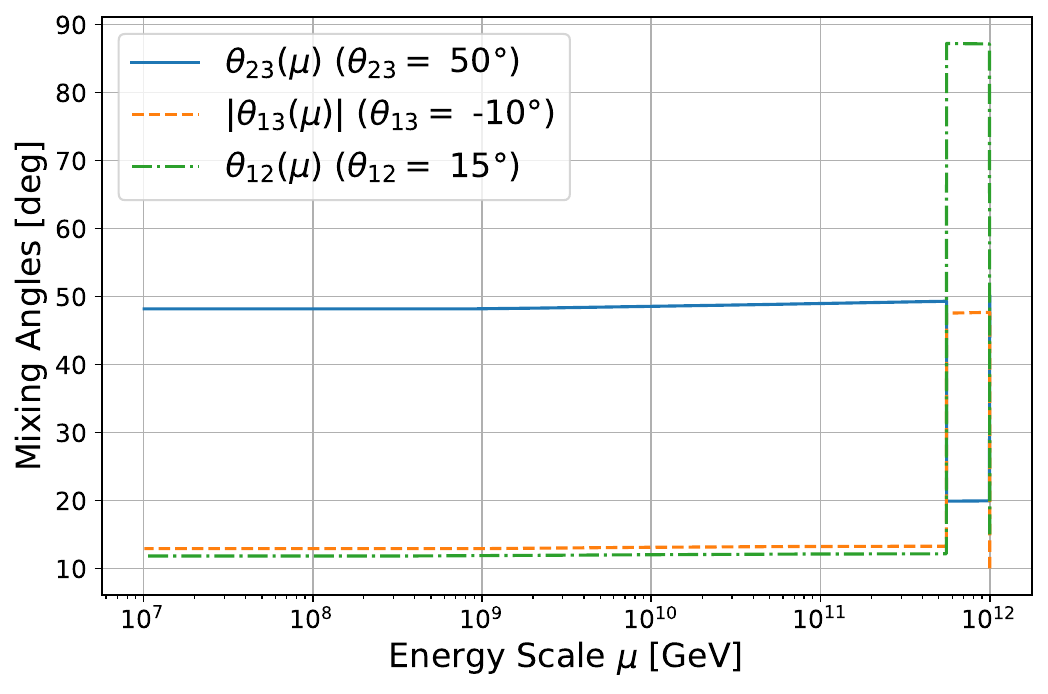}
    \caption{Running of the eigenvalues (left panel) and the mixing angles (right panel) for a scenario where at the cut-off scale $\kappa_3=1$ (in arbitrary units), $\kappa_1=\kappa_2=\kappa_3/100$, $\theta_{12}=15^\circ$, $|\theta_{13}|=10^\circ$, $\theta_{23} = 50^\circ$, and when an $L_\mu-L_\tau$ gauge boson has a mass $M_{Z'}=10^{9}$ GeV and the gauge coupling is $g'=1$.}
\label{fig:k1eqk2eqk3_100}
\end{figure}

We show in figure~\ref{fig:k1eqk2eqk3_100} the running of the modulus of the eigenvalues (left panel) and the mixing angles (right panel) for the specific case where $M_1=10^{12}$ GeV, $M_{Z'}=10^{9}$ GeV and $g'=1$, neglecting the contribution from $\alpha$. The value of $M_{Z'}$ only enters as the scale below which the running stops, such that for larger $M_{Z'}$, the resulting values can be directly read off at the respective scale. As initial conditions of the eigenvalues at the cut-off we choose $\kappa_3=1$ (in arbitrary units), $\kappa_1=\kappa_2=\kappa_3/100$, and the mixing angles $\theta_{23}=50^\circ$, $|\theta_{13}|=10^\circ$, $\theta_{12}=15^\circ$ (the value of $\theta_{12}$ is irrelevant, due to the degeneracy of $\kappa_1$ and $\kappa_2$). 

As shown in the plot, the degenerate eigenvalues are split according to eq.~\eqref{eq:mass-3gen-normal}, and the mixing angles including the initially unphysical $\theta_{12}$ are pushed to a fixed point satisfying eq.~\eqref{eq:angle-3gen-normal}. Below $M_1$, $\theta_{12}$ becomes physical due to the splitting of $\kappa_1$ and $\kappa_2$. Note that enforcing $|\kappa_1(\mu)| \leq |\kappa_2(\mu)|$ for all $\mu$ leads to the eigenvalues being swapped throughout the running, due to $({\rm d}\kappa_1/{\rm d}t) |_{M_1} \simeq 0,\, ({\rm d}\kappa_2/{\rm d}t) |_{M_1} < 0$. This causes the jumps in the mixing angles at the level crossing points, at $10^{12}$ GeV and $\sim 5 \cdot 10^{11}$ GeV. This can be verified by introducing a small splitting between $\kappa_1$ and $\kappa_2$, or setting them to zero as we will discuss shortly. The latter case also shows the initial jump due to the fixed point condition in isolation, while it is masked in this case by the reordering of the eigenvalues.

In the special case when $\kappa_1 = \kappa_2 = 0$ at the cut-off scale, then $W_{31}=0$ or $W_{32}=0$ (cf. eqs.~\eqref{eq:T_RGE}, \eqref{eq:degen_quasifixed_val}), implying that one of the eigenvalues must remain zero at leading order. Following again the convention that $|\kappa_1(\mu)|\leq |\kappa_2(\mu)| $ for $\mu < M_1$, one obtains the prediction $W_{31}=0$, as well as the eigenvalues
\begin{align}
\kappa_1|_{M_{Z'}}&= 0
\nonumber \\
\kappa_2|_{M_{Z'}}&\simeq 
-b W_{32}^2 \kappa_3 \nonumber \\
\kappa_3|_{M_{Z'}}&\simeq  \kappa_3.
\label{eq:mass-3gen-0}
\end{align}

On the other hand, $\kappa_1$ is not protected by any symmetry and could be generated by higher-order effects, in particular, by the running of the mixing angles and $\kappa_2$ from $M_1$ to $M_{Z'}$. 
Indeed, using the exact solution eq.~\eqref{eq:exact}, and the invariants  $\kappa_1 \kappa_2+\kappa_1 \kappa_3+\kappa_2\kappa_3=1/2\big({\rm tr}(\kappa)^2-{\rm tr}(\kappa^2)\big)$, $\kappa_1\kappa_2\kappa_3={\rm det}(\kappa)$, one obtains
\begin{align}
\kappa_1|_{M_{Z'}}\simeq\frac{2{\rm det}(\kappa)}{{\rm tr}(\kappa)^2-{\rm tr}(\kappa^2)}\Big|_{M_{Z'}}\simeq 2b^2 \kappa_3  \frac{U_{31}^2U_{32}^2 U_{33}^2}{4 U_{32}^2U_{33}^2+U_{31}^2 U_{32}^2 + U_{31}^2 U_{33}^2}\;.
\end{align}

This expression is  $\propto b^2 \sim 1/(16\pi^2)^2\log^2(M_1/M_{Z'})$, and could receive additional contributions from two-loop effects,  which are of order $1/(16\pi^2)^2\log(M_1/M_{Z'})$. Since $\kappa_1$ is difficult to measure experimentally, we will not pursue in this paper a full two-loop calculation, and we will limit ourselves to calculate the order of magnitude of the radiatively generated $\kappa_1$.

\begin{figure}[t!]
    \centering
    \includegraphics[width=0.45\textwidth]{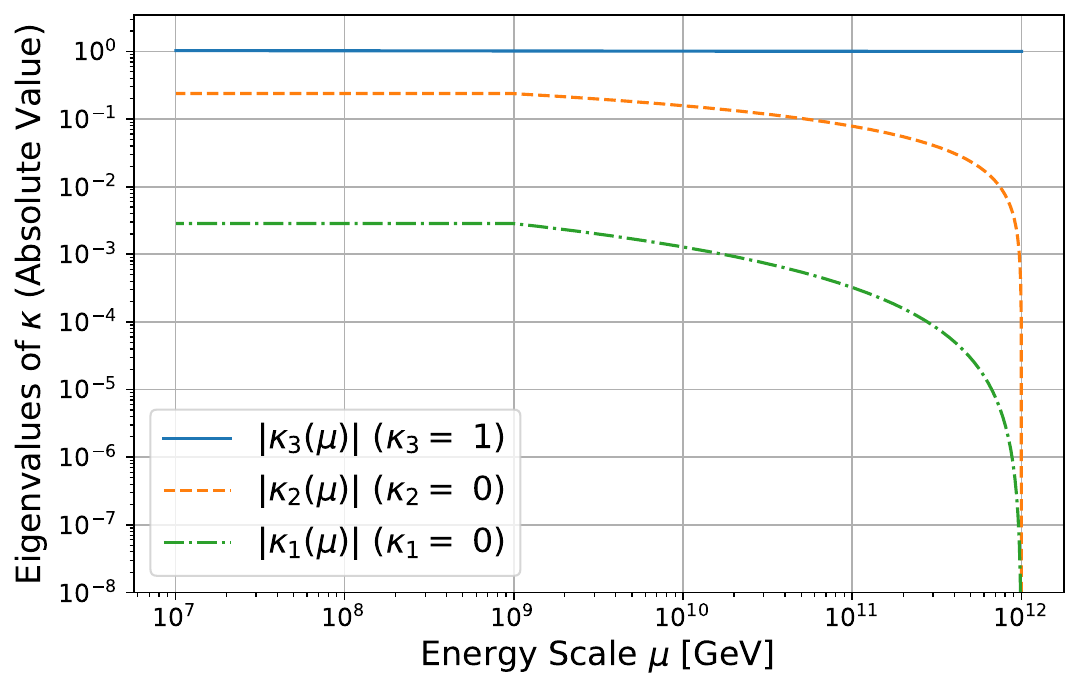}
    \includegraphics[width=0.45\textwidth]{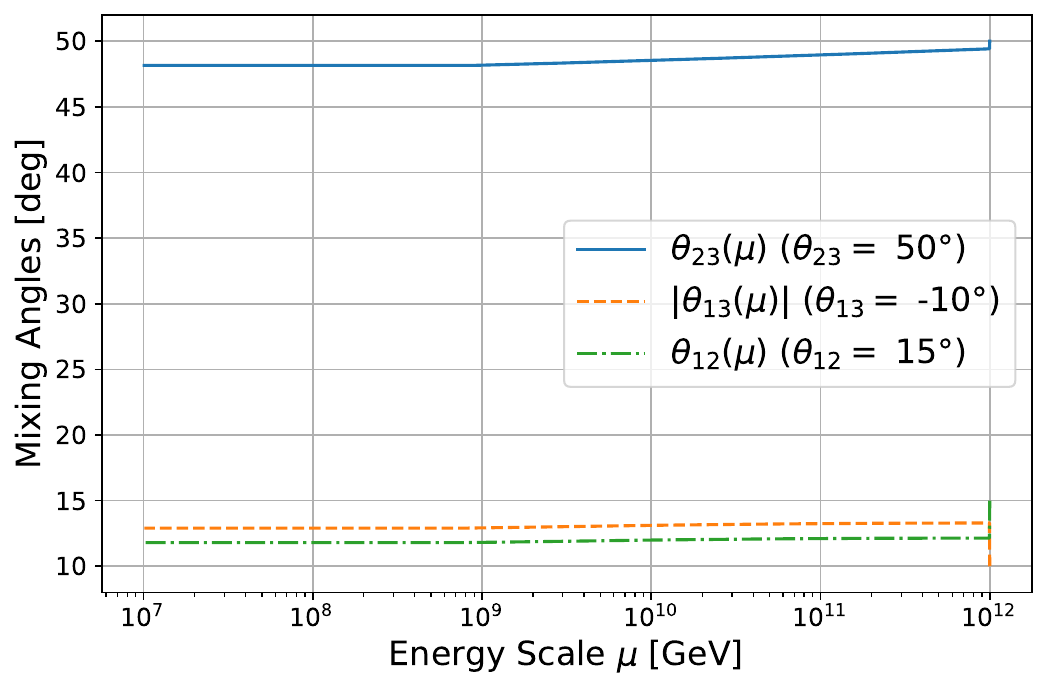}
    \caption{Running of the eigenvalues (left panel) and the mixing angles (right panel) for a scenario where at the cut-off scale $\kappa_3=1$ (in arbitrary units), $\kappa_1=\kappa_2=0$, $\theta_{12}= 15^\circ$, $|\theta_{13}|=10^\circ$, $\theta_{23} = 50^\circ$, and when an $L_\mu-L_\tau$ gauge boson has a mass $M_{Z'}=10^{9}$ GeV and the gauge coupling is $g'=1$.}
  \label{fig:k1eqk2eq0}  
\end{figure}

The numerical results for this case are shown in figure~\ref{fig:k1eqk2eq0}. As one can see by the small jumps in the mixing angles, they are instantaneously pushed to the fixed point of eq.~\eqref{eq:angle-3gen-normal}, $W_{31} = 0$. If there were a small splitting between $\kappa_1$ and $\kappa_2$, the fixed point would be reached smoothly, whereas the exact degeneracy forces $W_{31} = 0$ to be exact as well, leading to the jump. We furthermore observe that since the mixing angles do not run significantly, the fixed point persists approximately even in the infrared, where the degeneracy of $\kappa_1, \kappa_2$ is broken.

For the experimental central values of the atmospheric and solar mixing angles, $\theta_{12}\simeq 34^\circ$, $\theta_{23}\simeq 48^\circ$, the predicted value for the reactor mixing angle is $|\theta_{13}|\simeq 23^\circ$, which is far from the experimental result $\theta_{13}=(8.52^{+0.11}_{-0.11})^\circ$ ~\cite{Esteban:2024eli}. It is noteworthy that in the presence of complex couplings the quasifixed point condition is ${\rm Re}\big[W_{31} W_{32}\big]=0$, which involves phases. Therefore, by adjusting the Majorana and CP phases it is possible to reproduce the quasifixed point condition, even for the measured values of the mixing angles. A similar statement holds for ${\rm Im}\big[W_{31} W_{32}\big]=0$ in the case of degenerate eigenvalues with opposite sign (cf. eq.~\eqref{eq:T_RGE}).
A detailed analysis of the complex case will be presented elsewhere. 

\begin{figure}[t!]
    \centering
    \includegraphics[width=0.45\textwidth]{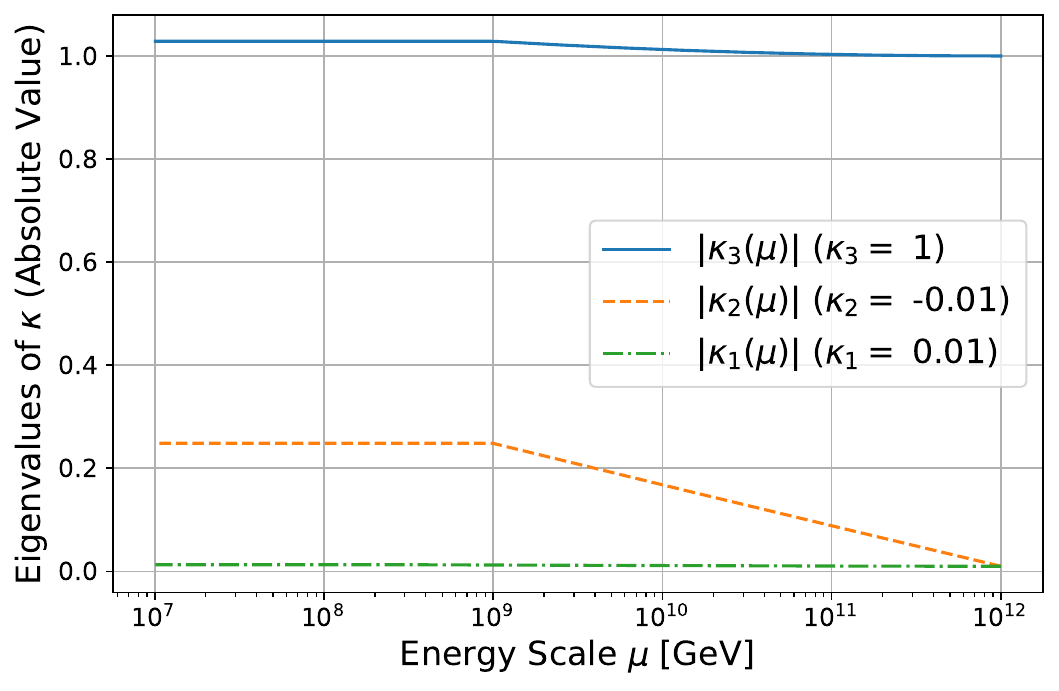}~~
    \includegraphics[width=0.45\textwidth]{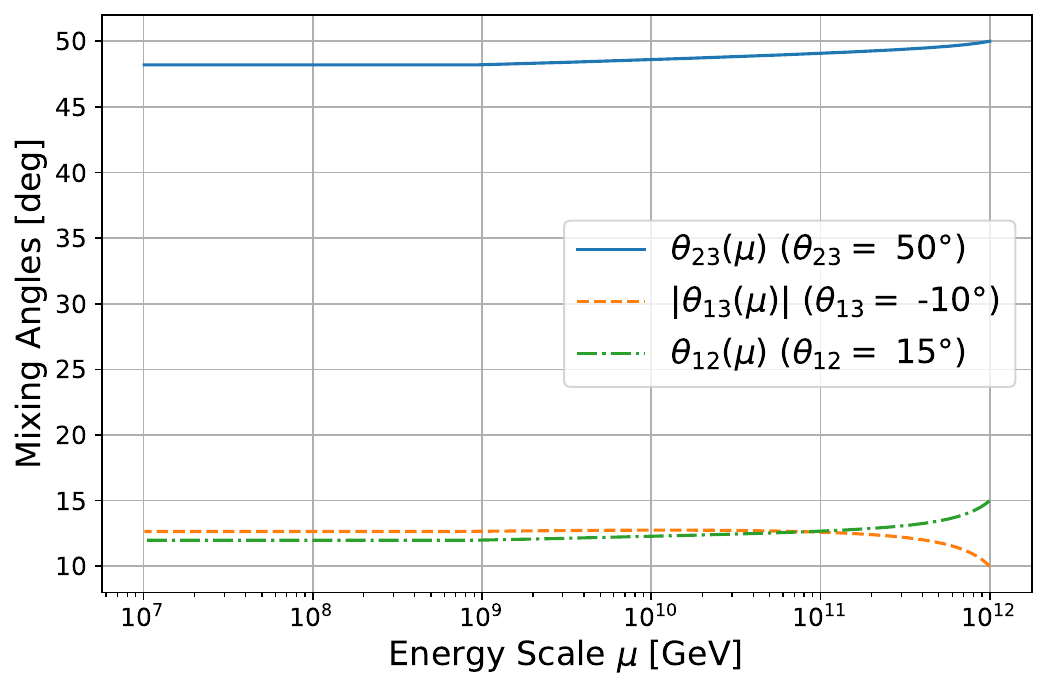}
    \caption{Same as figure~\ref{fig:k1eqk2eqk3_100}, but for $\kappa_1=-\kappa_2$.}
\label{fig:-k1eqk2eqk3_100}
\end{figure}

On the other hand, if $\kappa_1$ and $\kappa_2$ are degenerate in absolute value but have opposite CP phases, namely $\kappa_1 = -\kappa_2$, then
\begin{align}
\big|\kappa_1|_{M_{Z'}}\big|&\simeq {\min_{i=1,2}}\Big\{\big|
\big[(-1)^i a+b W_{1i}^2-b W_{2i}^2\big]\kappa_2-b W_{3i}^2 \kappa_3\big|\Big\}
\nonumber \\
\big|\kappa_2|_{M_{Z'}}\big|&\simeq {\max_{i=1,2}}\Big\{
\big|\big[(-1)^i a+b W_{1i}^2-b W_{2i}^2\big]\kappa_2-b W_{3i}^2 \kappa_3\big|\Big\}\nonumber \\
\big|\kappa_3|_{M_{Z'}}\big|&\simeq  \big|\kappa_3\big|,
\end{align}
with no fixed point for the mixing angles. We show this scenario in figure~\ref{fig:-k1eqk2eqk3_100}. Since there is no fixed point condition in this case, the mixing angles run smoothly even for exactly degenerate $|\kappa_1|, |\kappa_2|$. Note that fixed points for the mixing angles may be obtained if the mixing matrix $U$ contains non-zero phases, as previously mentioned.

\subsubsection{Inverted mass hierarchy: \texorpdfstring{$|\kappa_3|\ll |\kappa_1|\simeq |\kappa_2|$}{|k3| << |k1| ~ |k2|}}

Similarly to the normal hierarchy scenario, the lightest neutrino mass (in this case proportional to $\kappa_3$) can receive at the one-loop level significant contributions from the heavier eigenvalues ($\kappa_1$ and $\kappa_2$):
\begin{align}
\kappa_1|_{M_{Z'}}&\simeq \big(a - b W_{11}^2\big)\kappa_1- b W_{21}^2  \kappa_2  \nonumber \\
\kappa_2|_{M_{Z'}}&\simeq -b W_{12}^2\kappa_1+ \big(a - b W_{22}^2\big)\kappa_2 \nonumber \\
\kappa_3|_{M_{Z'}}&\simeq -b W_{13}^2\kappa_1- b W_{23}^2\kappa_2.
\end{align}
This generates a hierarchy between the smallest and the largest eigenvalues which approximately reads:
\begin{align}
\frac{\kappa_3}{\kappa_2}
 \simeq -\frac{6 g'^2}{16\pi^2}
(U_{32}U_{22}- U_{33}U_{23})^2\log\left(\frac{M_1}{M_{Z'}}\right).
\end{align}

\begin{figure}[t!]
    \centering
    \includegraphics[width=0.45\textwidth]{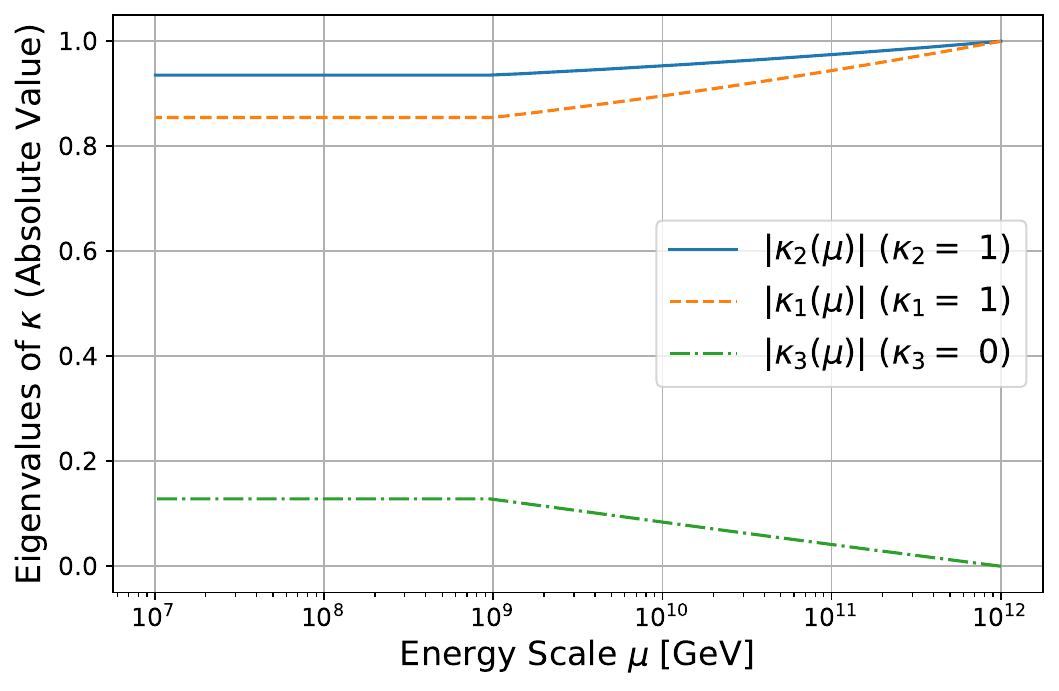}~~
    \includegraphics[width=0.45\textwidth]{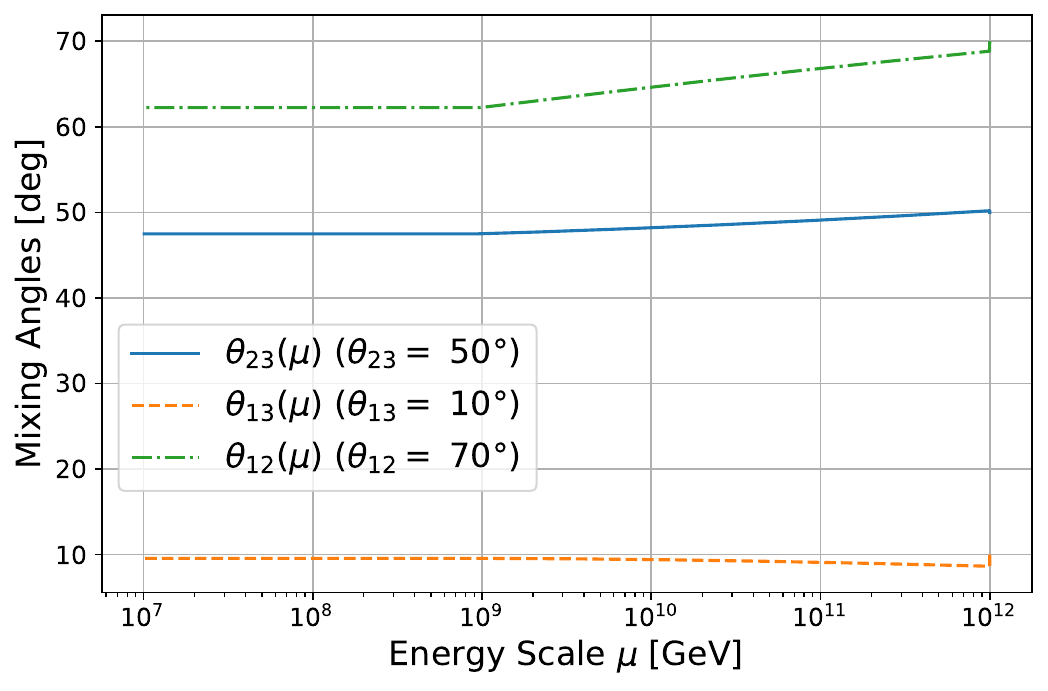}
    \caption{Running of the masses (left panel) and the mixing angles (right panel) for a scenario where at the cut-off scale $\kappa_1=\kappa_2=1$ (in arbitrary units), $\kappa_3=0$, $\theta_{12}=70^\circ$, $\theta_{13}=10^\circ$, $\theta_{23} = 50^\circ$ when a $L_\mu-L_\tau$ gauge boson has a mass $M_{Z'}=10^{9}$ GeV and the gauge coupling is $g'=1$.}
\label{fig:k1eqk2ggk3}
\end{figure}

Of particular interest is the case where $\kappa_3\ll \kappa_1=\kappa_2$, which could split the two largest eigenvalues. In this case,  the  mass eigenstates read:
\begin{align}
\big|\kappa_1|_{M_{Z'}}\big|&\simeq {\min_{i=1,2}}\Big\{\big|
\big(a-b W_{1i}^2-b W_{2i}^2\big)\kappa_2\big|\Big\} \nonumber \\
\big|\kappa_2|_{M_{Z'}}\big|&\simeq {\max_{i=1,2}}\Big\{\big|
\big(a-b W_{1i}^2-b W_{2i}^2\big)\kappa_2\big|\Big\} \nonumber \\
\big|\kappa_3|_{M_{Z'}}\big|&\simeq   \big|- b \big(W_{13}^2 + W_{23}^2\big) \kappa_2\big| ,
\label{eq:mass-3gen-inv}
\end{align}
while the mixing angles then quickly reach a fixed point corresponding to 
$W_{11} W_{12} + W_{21} W_{22}=0$. This case is illustrated in figure~\ref{fig:k1eqk2ggk3}, which assumes $\kappa_1=\kappa_2$, $\kappa_3=0$, and for the mixing angles $\theta_{12} = 70^\circ$, $\theta_{13} = 10^\circ$, $\theta_{23} = 50^\circ$. Again, we observe a small jump in the mixing angles as they are instantaneously pushed to the fixed point due to the exact degeneracy of $\kappa_1, \kappa_2$.

In general, $\kappa_1$ and $\kappa_2$ are split, generating a solar mass splitting approximately given by:
\begin{align}
(m_2^2-m_1^2)|_{M_{Z'}}& \simeq b\, m_2^2\,\big|W_{11}^2-W_{22}^2\big|\nonumber \\
&\simeq \frac{6 g'^2}{16\pi^2}m_2^2\big|(U_{12}^2 - U_{13}^2)^2 - (U_{22}^2 - U_{23}^2)^2\big| \log\left(\frac{M_1}{M_{Z'}}\right),
\end{align}
which is in the ballpark of the measured values when $g'\simeq 1$. In the real case, the fixed point is $W_{11} W_{12} + W_{21} W_{22} = W_{12}(W_{11} + W_{22}) = 0$, which can be satisfied by $W_{11} = -W_{22}=0$ or $W_{12}=0$. For mixing angles satisfying $W_{11} = -W_{22}=0$, the eigenvalues remain degenerate at $\mathcal{O}(b)$, with a splitting induced at $\mathcal{O}(b^2) \sim 1/(16\pi^2)^2\log^2(M_1/M_{Z'})$. On the other hand, when the fixed point is satisfied through $W_{12}=0$, $W_{11}$ and $W_{22}$ are independent and the eigenvalues are split at $\mathcal{O}(b)$, inducing a potentially larger splitting depending on the initial conditions. We show this case in figure~\ref{fig:k1eqk2ggk3}. 

\begin{figure}[t!]
    \centering
    \includegraphics[width=0.45\textwidth]{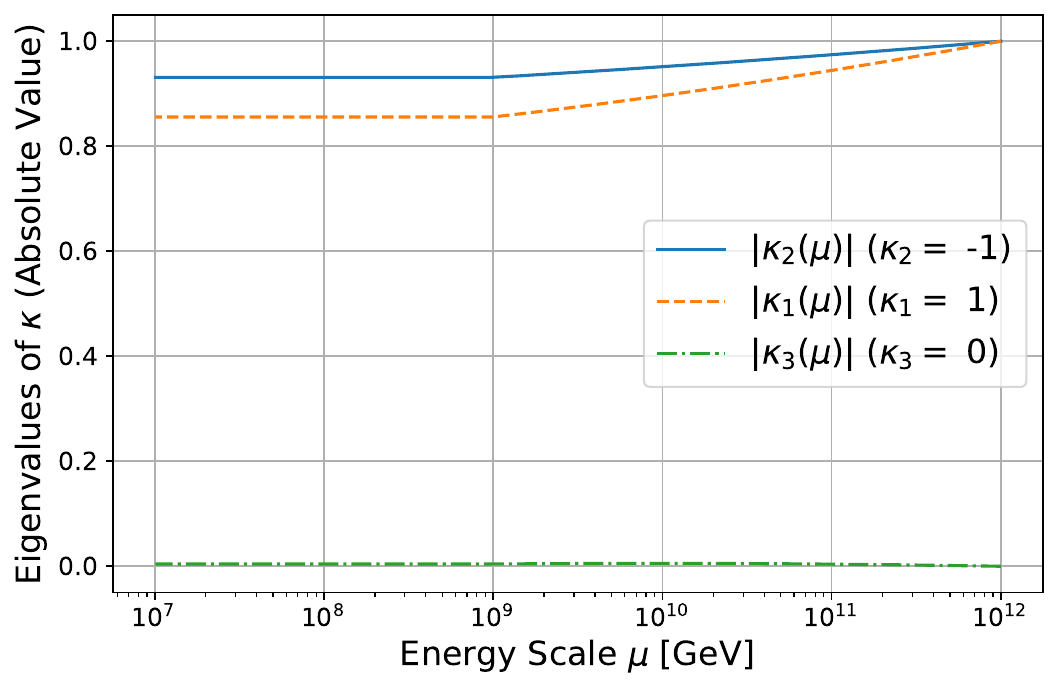}~~
    \includegraphics[width=0.45\textwidth]{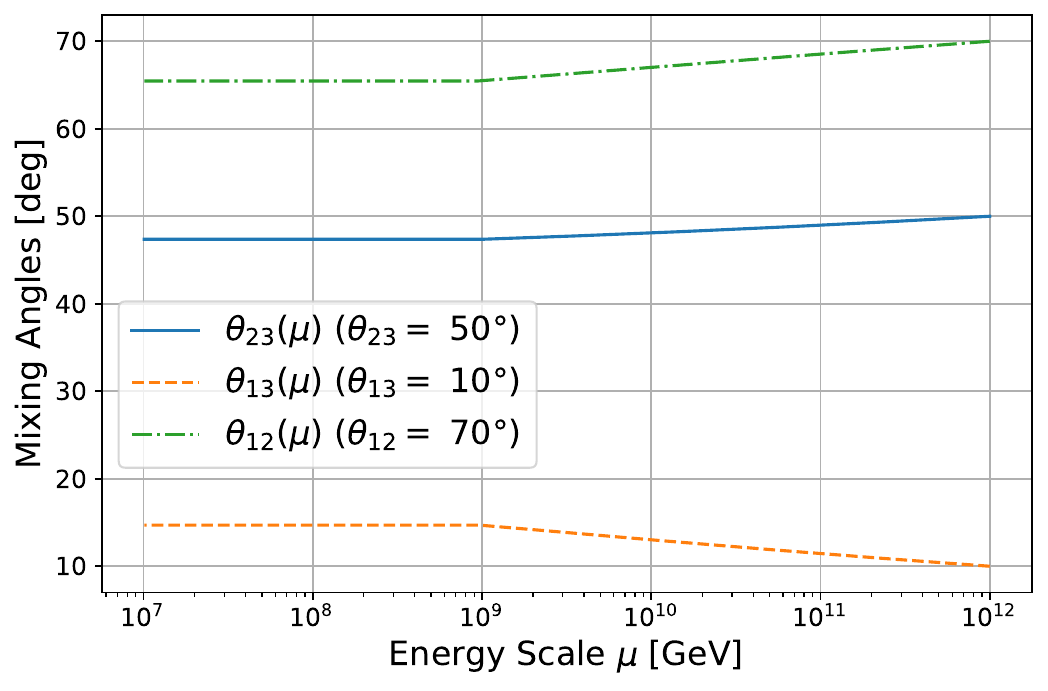}
    \caption{Same as figure~\ref{fig:k1eqk2ggk3}, but for $\kappa_1=-\kappa_2$.}
\label{fig:-k1eqk2ggk3}
\end{figure}

Considering again the case where $\kappa_1=-\kappa_2$, we find 
\begin{align}
\big|\kappa_1|_{M_{Z'}}\big|&\simeq {\min_{i=1,2}}\Big\{\big|
\big[(-1)^i a+b W_{1i}^2-b W_{2i}^2\big]\kappa_2\big|\Big\} \nonumber \\
\big|\kappa_2|_{M_{Z'}}\big|&\simeq {\max_{i=1,2}}\Big\{\big|
\big[(-1)^i a+b W_{1i}^2-b W_{2i}^2\big]\kappa_2\big|\Big\} \nonumber \\
\big|\kappa_3|_{M_{Z'}}\big|&\simeq \big| b \big(W_{13}^2 - W_{23}^2\big)  \kappa_2 \big|,
\label{eq:mass-3gen-inv-minus}
\end{align}
and again no fixed point for the mixing angles.  We show this case in figure~\ref{fig:-k1eqk2ggk3}. Note that $\kappa_3|_{M_{Z'}}$ is small but non-zero due to a partial cancellation between $W_{13}^2$ and $W_{23}^2$ as shown in eq.~\eqref{eq:mass-3gen-inv-minus}, in contrast to the case of eq.~\eqref{eq:mass-3gen-inv}.

\section{Conclusions}
\label{sec:conclusions}

We have considered a scenario where the Standard Model Lagrangian is extended with a dimension-5 lepton-number-breaking operator (the Weinberg operator), and where the particle content is extended by a gauge boson with flavor-dependent couplings. Specifically, we have considered a $\textrm{U}(1)_{L_\mu-L_\tau}$ extension, although our main conclusions can also be applied for other Abelian or non-Abelian gauge symmetries. This scenario can arise (for example) in models with right-handed neutrinos and scalars that are Standard Model singlets charged under the $\textrm{U}(1)_{L_\mu-L_\tau}$ symmetry, and where the scale of spontaneous symmetry breaking of the flavor-dependent symmetry is much larger than the scale of the right-handed neutrino mass terms. In this case, the pole masses of the right-handed neutrinos are related to the mass of the gauge boson associated with the spontaneous symmetry breaking. 

We have found that there are wide regions of parameter space where the flavored gauge boson is lighter than the lightest right-handed neutrino, thus leading to an effective field theory with a Weinberg operator and a massive, flavored gauge boson. The Yukawa couplings of the right-handed neutrinos to the scalars and the left-handed leptons remain unspecified in our analysis, hence the eigenvalues and eigenvectors of the Weinberg operator can take in general any value, allowing mild hierarchies, large hierarchies, or even very small parameters.  

In scenarios with heavy right-handed neutrinos, the Weinberg operator is generated at a high energy scale, making it necessary to properly include quantum effects to connect the model parameters at the cut-off of the effective theory to their values at low energies measured in experiments. The leading quantum corrections in this case are encoded in the renormalization group equations (RGEs). In this work, we have calculated for the first time the RGE of the dimension-five Weinberg operator in the presence of flavor-dependent gauge interactions, and we have discussed the most relevant features and phenomenology of the RGE effects.

We have shown that  a flavor-dependent gauge boson coupling to the left-handed leptons can increase the rank of the neutrino mass matrix at the one-loop level. This is in stark contrast to the well-studied scenario with just the particle content of the Standard Model, where one-loop effects do not change the rank of the mass matrix, and only two-loop effects can increase the rank. We have also considered several scenarios with degenerate neutrinos, and we have determined the quasifixed points in the infrared that arise for the elements in the leptonic mixing matrix, as well as the mass splittings that are generated due to quantum effects, naturally leading to mass hierarchies consistent with experiment.

\subsection*{Acknowledgments}

This work is supported by the Collaborative Research Center SFB1258, by the Deutsche Forschungsgemeinschaft (DFG, German Research Foundation) under Germany's Excellence Strategy - EXC-2094 - 390783311, and by JSPS KAKENHI Grant Number 22K21350.

\appendix

\section{Non-Abelian generalization} \label{appendix:non-abelian}

The structure presented in this paper for an Abelian, flavored gauge symmetry can be straightforwardly generalized to non-Abelian gauge groups. In the neutral basis ({\it c.f.} $W^1, W^2, W^3$ for $\textrm{SU}(2)_L$), every gauge boson gives rise to one $G^T \kappa G$ term in the RGE, where each $G$ is proportional to the respective generator. In the charged basis ({\it c.f.} $W^\pm, W^3$ for $\textrm{SU}(2)_L$), mixed terms between the conjugate generators arise. The resulting RGE takes the form

\begin{equation}
    16 \pi^2 \beta_\kappa\, = \alpha\, \kappa + P^T\, \kappa + \kappa\, P + G^T\, \kappa\, G + \frac{1}{2} \big(G_+^T\, \kappa\, G_- + G_-^T\, \kappa\, G_+\big).
\end{equation}
Here, we show the simplest case of one neutral and two conjugate, flavor-charged gauge bosons. For larger gauge groups, there is a corresponding sum over the respective generators.

Following the same calculation as in the Abelian case, we obtain for the RGEs
\begin{equation}
\begin{aligned}
    16 \pi^2 \frac{{\rm d} \kappa_i}{{\rm d} t} = \alpha\, \kappa_i + 2 \tilde{P}_{ii} \kappa_i &+ \sum_{k = 1}^n {\rm Re}\big[\tilde{G}_{ki}^2 + \tilde{G}_{+, ki} \tilde{G}_{-, ki}\big] \kappa_k, 
    \end{aligned}
\end{equation}
with $\tilde{P} = U P U^\dagger, \tilde{G}_{(\pm)} = U G_{(\pm)} U^\dagger$, and $n$ the number of lepton generations. For the RGE of the mixing matrix, ${\rm d}U/{\rm d}t = T U$, we obtain
 \begin{equation}
    16 \pi^2 T_{ij} = \begin{cases}
    \begin{aligned}
    i \sum_{k = 1}^n {\rm Im}\big[\tilde{G}_{ki}^2 + \tilde{G}_{+, ki} \tilde{G}_{-, ki}\big] \frac{\kappa_k}{2\, \kappa_i} \end{aligned} 
    &\text{for $i = j$,}
    \\[15pt]
    \begin{aligned}
    &{\rm Re} [\tilde{P}_{ij}] \frac{\kappa_i + \kappa_j}{\kappa_i - \kappa_j} +  \sum_{k = 1}^n {\rm Re} \Big[\tilde{G}_{ki}\, \tilde{G}_{kj} + \frac{1}{2} \big(\tilde{G}_{+, ki} \tilde{G}_{-, kj} + \tilde{G}_{-, ki} \tilde{G}_{+, kj}\big)\Big]\times \\[5pt]
    &\times \frac{\kappa_k}{\kappa_i - \kappa_j} + i \bigg({\rm Im} \big[\tilde{P}_{ij}\big] \frac{\kappa_i- \kappa_j}{\kappa_i + \kappa_j} + \\[5pt]
    &+ \sum_{k = 1}^n {\rm Im} \Big[\tilde{G}_{ki}\, \tilde{G}_{kj} + \frac{1}{2} \big(\tilde{G}_{+, ki} \tilde{G}_{-, kj} + \tilde{G}_{-, ki} \tilde{G}_{+, kj}\big)\Big] \frac{\kappa_k}{\kappa_i + \kappa_j}\bigg)
    \end{aligned} 
    &\text{for $i \neq j$.}
    \end{cases}
\end{equation}

The explicit forms of $G_\pm$ are analogous to the Abelian case and given by

\begin{equation}
    G_\pm = \sqrt{6} g' Q'_\pm,
\end{equation}
where $Q'_\pm$ are the generators of the charged gauge bosons $W^a_\mu$ defined such that the neutral and charged bases as they would appear in the Lagrangian are related by
\begin{equation}
    W^a_\mu Q'_a = W^1_\mu Q'_1 + W^2_\mu Q'_2 + W^3_\mu Q'_3 = \frac{1}{\sqrt{2}} \big(W^+_\mu Q'_+ + W^-_\mu Q'_-\big) + W^3_\mu Q'_3. 
\end{equation}

We also note that if the lepton doublets are in the fundamental representation of an $\textrm{SU}(n)$ flavor gauge group, the generators fulfill the completeness relation (sum over $a$ implicit)

\begin{equation}
    T^a_{ij} T^a_{k\ell} = \frac{1}{2}\left(\delta_{i\ell}\delta_{jk} - \frac{1}{n}\delta_{ij}\delta_{k\ell}\right),
\end{equation}
see, e.g., \cite{Haber:2019sgz}. Based on this relation, we can derive for the contribution to $\beta_\kappa$

\begin{equation}
    \beta_{\kappa, j\ell} \supset \big[G_a^T \kappa G_a\big]_{j\ell} \propto \kappa_{ik}\,  Q'_{a, ij} Q'_{a, k\ell} = \kappa_{ik}\, \frac{1}{2}\left(\delta_{i\ell}\delta_{jk} - \frac{1}{n}\delta_{ij}\delta_{k\ell}\right) = \frac{n - 1}{2n} \kappa_{j\ell},
\end{equation}
where we used that $\kappa$ is symmetric in the last step. Since this contribution is proportional to $\kappa$ itself, the rank of $\kappa$ will not be raised when all $\textrm{SU}(n)$ gauge bosons contribute to $\beta_\kappa$ with the lepton doublets in the fundamental representation.

\clearpage
\printbibliography

@article{Ibarra:2024tpt,
    author = "Ibarra, Alejandro and Leister, Nicholas and Zhang, Di",
    title = "{Complete two-loop renormalization group equation of the Weinberg operator}",
    eprint = "2411.08011",
    archivePrefix = "arXiv",
    primaryClass = "hep-ph",
    reportNumber = "MITP-24-081",
    doi = "10.1007/JHEP03(2025)214",
    journal = "JHEP",
    volume = "03",
    pages = "214",
    year = "2025"
}

@article{Weinberg:1979sa,
    author = "Weinberg, Steven",
    title = "{Baryon and Lepton Nonconserving Processes}",
    reportNumber = "HUTP-79-A050",
    doi = "10.1103/PhysRevLett.43.1566",
    journal = "Phys. Rev. Lett.",
    volume = "43",
    pages = "1566--1570",
    year = "1979"
}

@article{Chankowski:1999xc,
    author = "Chankowski, Piotr H. and Krolikowski, Wojciech and Pokorski, Stefan",
    title = "{Fixed points in the evolution of neutrino mixings}",
    eprint = "hep-ph/9910231",
    archivePrefix = "arXiv",
    reportNumber = "CERN-TH-99-269, IFT-99-22",
    doi = "10.1016/S0370-2693(99)01465-3",
    journal = "Phys. Lett. B",
    volume = "473",
    pages = "109--117",
    year = "2000"
}

@article{Davidson:2006tg,
    author = "Davidson, Sacha and Isidori, Gino and Strumia, Alessandro",
    title = "{The Smallest neutrino mass}",
    eprint = "hep-ph/0611389",
    archivePrefix = "arXiv",
    reportNumber = "IFUP-TH-06-23",
    doi = "10.1016/j.physletb.2007.01.015",
    journal = "Phys. Lett. B",
    volume = "646",
    pages = "100--104",
    year = "2007"
}

@article{Ibarra:2018dib,
    author = "Ibarra, Alejandro and Strobl, Patrick and Toma, Takashi",
    title = "{Neutrino masses from Planck-scale lepton number breaking}",
    eprint = "1802.09997",
    archivePrefix = "arXiv",
    primaryClass = "hep-ph",
    reportNumber = "TUM-HEP-1133-18, KIAS-P18015",
    doi = "10.1103/PhysRevLett.122.081803",
    journal = "Phys. Rev. Lett.",
    volume = "122",
    number = "8",
    pages = "081803",
    year = "2019"
}

@article{Ibarra:2020eia,
    author = "Ibarra, Alejandro and Strobl, Patrick and Toma, Takashi",
    title = "{Two-loop renormalization group equations for right-handed neutrino masses and phenomenological implications}",
    eprint = "2006.13584",
    archivePrefix = "arXiv",
    primaryClass = "hep-ph",
    reportNumber = "TUM-HEP 1267/20",
    doi = "10.1103/PhysRevD.102.055011",
    journal = "Phys. Rev. D",
    volume = "102",
    number = "5",
    pages = "055011",
    year = "2020"
}

@article{Heeck:2011wj,
    author = "Heeck, Julian and Rodejohann, Werner",
    title = "{Gauged  $L_\mu  -  L_\tau$  Symmetry  at  the  Electroweak  Scale}",
    eprint = "1107.5238",
    archivePrefix = "arXiv",
    primaryClass = "hep-ph",
    doi = "10.1103/PhysRevD.84.075007",
    journal = "Phys. Rev. D",
    volume = "84",
    pages = "075007",
    year = "2011"
}

@article{He:1990pn,
    author = "He, X. G. and Joshi, Girish C. and Lew, H. and Volkas, R. R.",
    title = "{NEW Z-prime PHENOMENOLOGY}",
    reportNumber = "UM-P-90/42, OZ-P-90/16",
    doi = "10.1103/PhysRevD.43.R22",
    journal = "Phys. Rev. D",
    volume = "43",
    pages = "22--24",
    year = "1991"
}

@article{Casas:1999ac,
    author = "Casas, J. A. and Espinosa, J. R. and Ibarra, A. and Navarro, I.",
    title = "{Nearly degenerate neutrinos, supersymmetry and radiative corrections}",
    eprint = "hep-ph/9905381",
    archivePrefix = "arXiv",
    reportNumber = "IEM-FT-193-99, CERN-TH-99-142",
    doi = "10.1016/S0550-3213(99)00605-7",
    journal = "Nucl. Phys. B",
    volume = "569",
    pages = "82--106",
    year = "2000"
}

@article{Antusch:2001vn,
    author = {Antusch, Stefan and Drees, Manuel and Kersten, J{\"o}rn and Lindner, Manfred and Ratz, Michael},
    title = "{Neutrino mass operator renormalization in two Higgs doublet models and the MSSM}",
    eprint = "hep-ph/0110366",
    archivePrefix = "arXiv",
    reportNumber = "TUM-HEP-443-01",
    doi = "10.1016/S0370-2693(01)01414-9",
    journal = "Phys. Lett. B",
    volume = "525",
    pages = "130--134",
    year = "2002"
}

@article{Grimus:1999wm,
    author = "Grimus, W. and Neufeld, H.",
    title = "{Three neutrino mass spectrum from combining seesaw and radiative neutrino mass mechanisms}",
    eprint = "hep-ph/9911465",
    archivePrefix = "arXiv",
    reportNumber = "UWTHPH-1999-71, IFIC-99-84",
    doi = "10.1016/S0370-2693(00)00769-3",
    journal = "Phys. Lett. B",
    volume = "486",
    pages = "385--390",
    year = "2000"
}

@article{Grimus:2002nk,
    author = "Grimus, Walter and Lavoura, Luis",
    title = "{One-loop corrections to the seesaw mechanism in the multi-Higgs-doublet standard model}",
    eprint = "hep-ph/0207229",
    archivePrefix = "arXiv",
    reportNumber = "UWTHPH-2002-21",
    doi = "10.1016/S0370-2693(02)02672-2",
    journal = "Phys. Lett. B",
    volume = "546",
    pages = "86--95",
    year = "2002"
}

@article{Ibarra:2011gn,
    author = "Ibarra, Alejandro and Simonetto, Cristoforo",
    title = "{Understanding neutrino properties from decoupling right-handed neutrinos and extra Higgs doublets}",
    eprint = "1107.2386",
    archivePrefix = "arXiv",
    primaryClass = "hep-ph",
    doi = "10.1007/JHEP11(2011)022",
    journal = "JHEP",
    volume = "11",
    pages = "022",
    year = "2011"
}

@article{Antusch:2002ek,
    author = "Antusch, Stefan and Ratz, Michael",
    title = "{Supergraph techniques and two loop beta functions for renormalizable and nonrenormalizable operators}",
    eprint = "hep-ph/0203027",
    archivePrefix = "arXiv",
    reportNumber = "TUM-HEP-453-01",
    doi = "10.1088/1126-6708/2002/07/059",
    journal = "JHEP",
    volume = "07",
    pages = "059",
    year = "2002"
}

@article{Casas:1999tp,
    author = "Casas, J. A. and Espinosa, J. R. and Ibarra, A. and Navarro, I.",
    title = "{Naturalness of nearly degenerate neutrinos}",
    eprint = "hep-ph/9904395",
    archivePrefix = "arXiv",
    reportNumber = "IEM-FT-191-99, CERN-TH-99-103, IFT-UAM-CSIC-99-15",
    doi = "10.1016/S0550-3213(99)00383-1",
    journal = "Nucl. Phys. B",
    volume = "556",
    pages = "3--22",
    year = "1999"
}

@article{Casas:1999kc,
    author = "Casas, J. A. and Espinosa, J. R. and Ibarra, A. and Navarro, I.",
    title = "{Theoretical constraints on the vacuum oscillation solution to the solar neutrino problem}",
    eprint = "hep-ph/9906281",
    archivePrefix = "arXiv",
    reportNumber = "IEM-FT-195-99, CERN-TH-99-171, IFT-UAM-CSIC-99-23",
    doi = "10.1088/1126-6708/1999/09/015",
    journal = "JHEP",
    volume = "09",
    pages = "015",
    year = "1999"
}

@article{Haba:1999ca,
    author = "Haba, N. and Matsui, Y. and Okamura, N. and Sugiura, M.",
    title = "{Energy scale dependence of the lepton flavor mixing matrix}",
    eprint = "hep-ph/9904292",
    archivePrefix = "arXiv",
    reportNumber = "OHSTPY-HEP-T-99-010, DPNU-99-11, KEK-TH-620",
    doi = "10.1007/s100520050605",
    journal = "Eur. Phys. J. C",
    volume = "10",
    pages = "677--680",
    year = "1999"
}

@article{Esteban:2024eli,
    author = "Esteban, Ivan and Gonzalez-Garcia, M. C. and Maltoni, Michele and Martinez-Soler, Ivan and Pinheiro, Jo{\~a}o Paulo and Schwetz, Thomas",
    title = "{NuFit-6.0: updated global analysis of three-flavor neutrino oscillations}",
    eprint = "2410.05380",
    archivePrefix = "arXiv",
    primaryClass = "hep-ph",
    reportNumber = "IFT-UAM/CSIC-24-140, YITP-SB-2024-24, IPPP/24/64, IPPP/24/64, IFT-UAM/CSIC-24-140, YITP-SB-2024-24",
    doi = "10.1007/JHEP12(2024)216",
    journal = "JHEP",
    volume = "12",
    pages = "216",
    year = "2024"
}

@article{Denner:1992vza,
    author = "Denner, Ansgar and Eck, H. and Hahn, O. and Kublbeck, J.",
    title = "{Feynman rules for fermion number violating interactions}",
    reportNumber = "CERN-TH-6549-92",
    doi = "10.1016/0550-3213(92)90169-C",
    journal = "Nucl. Phys. B",
    volume = "387",
    pages = "467--481",
    year = "1992"
}

@article{Haber:2019sgz,
    author = "Haber, Howard E.",
    title = "{Useful relations among the generators in the defining and adjoint representations of SU(N)}",
    eprint = "1912.13302",
    archivePrefix = "arXiv",
    primaryClass = "math-ph",
    doi = "10.21468/SciPostPhysLectNotes.21",
    journal = "SciPost Phys. Lect. Notes",
    volume = "21",
    pages = "1",
    year = "2021"
}

@article{Ellis:1999my,
    author = "Ellis, John R. and Lola, Smaragda",
    title = "{Can neutrinos be degenerate in mass?}",
    eprint = "hep-ph/9904279",
    archivePrefix = "arXiv",
    reportNumber = "CERN-TH-99-87",
    doi = "10.1016/S0370-2693(99)00545-6",
    journal = "Phys. Lett. B",
    volume = "458",
    pages = "310--321",
    year = "1999"
}

@article{Antusch:2001ck,
    author = {Antusch, Stefan and Drees, Manuel and Kersten, J\"orn and Lindner, Manfred and Ratz, Michael},
    title = "{Neutrino mass operator renormalization revisited}",
    eprint = "hep-ph/0108005",
    archivePrefix = "arXiv",
    reportNumber = "TUM-HEP-424-01",
    doi = "10.1016/S0370-2693(01)01127-3",
    journal = "Phys. Lett. B",
    volume = "519",
    pages = "238--242",
    year = "2001"
}

@article{Babu:1993qv,
    author = "Babu, K. S. and Leung, Chung Ngoc and Pantaleone, James T.",
    title = "{Renormalization of the neutrino mass operator}",
    eprint = "hep-ph/9309223",
    archivePrefix = "arXiv",
    reportNumber = "IUHET-252, UDHEP-93-03, BA-93-44",
    doi = "10.1016/0370-2693(93)90801-N",
    journal = "Phys. Lett. B",
    volume = "319",
    pages = "191--198",
    year = "1993"
}

@article{Chankowski:1993tx,
    author = "Chankowski, Piotr H. and Pluciennik, Zbigniew",
    title = "{Renormalization group equations for seesaw neutrino masses}",
    eprint = "hep-ph/9306333",
    archivePrefix = "arXiv",
    reportNumber = "ZU-TH-20-93, DFPD-93-TH-44",
    doi = "10.1016/0370-2693(93)90330-K",
    journal = "Phys. Lett. B",
    volume = "316",
    pages = "312--317",
    year = "1993"
}

@article{Chankowski:2001hx,
    author = "Chankowski, Piotr H. and Wasowicz, Pawel",
    title = "{Low-energy threshold corrections to neutrino masses and mixing angles}",
    eprint = "hep-ph/0110237",
    archivePrefix = "arXiv",
    reportNumber = "IFT-01-28",
    doi = "10.1007/s100520100867",
    journal = "Eur. Phys. J. C",
    volume = "23",
    pages = "249--258",
    year = "2002"
}

@article{Casas:1999tg,
    author = "Casas, J. A. and Espinosa, J. R. and Ibarra, A. and Navarro, I.",
    title = "{General RG equations for physical neutrino parameters and their phenomenological implications}",
    eprint = "hep-ph/9910420",
    archivePrefix = "arXiv",
    reportNumber = "IEM-FT-196-99, IFT-UAM-CSIC-99-40, CERN-TH-99-315",
    doi = "10.1016/S0550-3213(99)00781-6",
    journal = "Nucl. Phys. B",
    volume = "573",
    pages = "652--684",
    year = "2000"
}

@article{Chankowski:2001mx,
    author = "Chankowski, Piotr H. and Pokorski, Stefan",
    title = "{Quantum corrections to neutrino masses and mixing angles}",
    eprint = "hep-ph/0110249",
    archivePrefix = "arXiv",
    reportNumber = "IFT-01-27",
    doi = "10.1142/S0217751X02006109",
    journal = "Int. J. Mod. Phys. A",
    volume = "17",
    pages = "575--614",
    year = "2002"
}

@article{Chankowski:2000fp,
    author = "Chankowski, Piotr H. and Ioannisian, A. and Pokorski, S. and Valle, J. W. F",
    title = "{Neutrino unification}",
    eprint = "hep-ph/0011150",
    archivePrefix = "arXiv",
    reportNumber = "FERMILAB-PUB-00-250-T, IFIC-00-68, IFT-00-26",
    doi = "10.1103/PhysRevLett.86.3488",
    journal = "Phys. Rev. Lett.",
    volume = "86",
    pages = "3488--3491",
    year = "2001"
}

\end{document}